# FOR BETTER OR FOR WORSE? A FRAMEWORK FOR CRITICAL ANALYSIS OF ICT4D FOR WOMEN

Abhipsa Pal, Indian Institute of Management Kozhikode, abhipsapal@iimk.ac.in

Rahul De', Indian Institute of Management Bangalore, rahul@iimb.ac.in

**Abstract:** Diffusion of ICTs provide possibilities for women empowerment by greater participation and enhanced gender-based digital equality. However, a critical analysis reveals that as ICT diffusion widens, there is a persistent threat of widening the gender-based digital divide and exposes women to online sexual abuses, predominantly in developing countries characterized by the gendered nature of the social structure. Instead of accepting ICT as the facilitator to women empowerment, in this paper, we develop a critical research framework for a gender-focused examination of ICT4D studies. Critical research methodology provides the appropriate ontology unveiling social realities through challenging the status quo and exposing the deeper societal inequalities. Using the critical research framework developed, we investigate past ICT4D initiatives and artifacts from literature and draw critical conclusions of its benefits and issues. This study would aid future ICT4D research to investigate areas of gender discrimination and understand the role of ICTs in a critical light.

## 1. INTRODUCTION

ICTs today hold potential for women empowerment in various ways including easier expression of their feelings through text messages (Barendregt, 2008), spreading awareness about women's rights on social media (Newsom & Lengel, 2012), and easier access to their finances through mobile banking (Kemal & Yan, 2015). The diffusion of mobile phones and the Internet have dramatically changed the ICT usage landscape for women, with a near equal number of women users, even in certain developing regions in Africa and South America (Gillwald, Milek, & Stork, 2010; Poushter, 2016). On the other hand, the variety of articles highlight the undeniably heroic role of ICTs in enabling women's voices and increased access to various services, advancing women's quality of life and social status (Ahmed, Islam, Hasan, & Rahman, 2006). However, on the other hand, a critical view of empowering technology is the dark side of ICT usage by women, the gender-based issues that gain prominence in the digital world. There is an evident threat of gender inequality to widen as ICT diffuses across geographies because even "ICT is not gender blind" (Ibrahim & Adamu, 2016). For instance, the gender divide in terms of smartphone accessibility is dangerously skewed with as low as only 17 percent of women mobile subscribers (Rathee, 2018), in India, with Kenya and Tunisia being close followers (Silver et al., 2019). Sexual abuses and harassment happen predominantly in social media and through mobile phone communications. This evidence shows that the gender-focused studies on ICT usage cannot ignore the deep socio-political issues and historical culture of the economies in question, which leads to persistent gender discrimination. In this paper, *we aim to critically examine the role of ICT in empowering women and investigate the persisting gender-based inequality through a review of the past ICT4D literature.*

Past literature has focused on the provision of equal access to digital technologies to men and women, thereby establishing how ICT4D aids in reducing the gaping divide (Akubue, 2000; Danjuma, Onimode, & Onche, 2015). However, field studies reveal gender-related challenges in terms of operationalization of ICT4D projects due to the present inequality of access (Potnis, 2014). Though gender-focused ICT studies largely identify ICTs in positive roles (Roberts, 2016; Seth et al., 2020), a rising number of studies now have raised the question of continued bias towards women even in the use and participation of ICTs (Moitra, Hassan, Mandal, Bhuiyan, & Ahmed, 2020; Potnis, 2014). Developing countries are often characterized by historic traditions that give rise to systematic gender bias, reinforced in the ICT usage scenario (Ibtasam et al., 2019). The gendered nature of societal problems of bias and abuse are often prominent in regions influenced by postcolonialism. Therefore, we argue that the ICT4Ds role should be critically analyzed since there is evidence of gender discrimination, in terms of ICT and social structures in these developing economies.

Critical research methodology aims to critically view and examine the status quo, the existing environment in which ICT is implemented, examining the orthodox traditions, historical social injustices, and deeply ingrained inequalities (De', Pal, Sethi, Reddy, & Chitre, 2017). In this paper, we first develop a framework for critical research for gender studies and then analyze ICT4D initiatives using data from past literature. We aim to undergo a literature review to analyze the state of gender-focused studies in ICT4D literature and analyze the articles critically to identify the missing context of discriminatory roles of ICT. Based on the findings, we suggest a framework for the reference of future studies that reveals the various dominating themes of gender studies within the ICT4D domain, and how critical research methodology can be beneficial.

The following part of the paper includes the motivation, followed by the theoretical background, and the methodology. The preliminary results are given with discussion concluding the paper.

## 2. MOTIVATION: ICT4D FOR WOMEN

*"Feminism is a social movement whose basic goal is equality between women and men. In many times and places in the past, people have insisted that women and men have similar capabilities and have tried to better the social position of all women…"*-(Lorber, 2001)

ICT4D promises eradication of gender inequality with greater freedom of expression and increased participation (Roberts, 2016). Interesting studies have reported that women spend more time on their mobile phones than men (Lemish & Cohen, 2005) and social media helped women protesters act at par with men (Newsom & Lengel, 2012). The variety of articles highlights the undeniably heroic role of ICTs in enabling women's voices and increased access to various services, advancing women's quality of life and social status (Ahmed et al., 2006).

In contrast, historically, information technology has reinforced gender discrimination, reported as early as the 1900s. Fewer women have exposure and literacy to computer usage due to distinctive absence of encouragement and support as their male counterparts, and face barriers to entry to Internet access just as a result of male dominance (Reisman, 1990; Spender, 1997). A critical view of empowering technology is the dark side of the ICT usage by women, the gender-based issues that gain prominence in the digital world. For instance, social media and mobile devices, while enabling women's rights activists and women protesters to express their opinions, give rise to a disturbing phenomenon of sexual harassment and gender-based abuses (Andalibi, Haimson, De Choudhury, & Forte, 2016; Delisle et al., 2019; Global Fund for Women, 2015). Across online platforms, women are regular victims of digital sexual crimes like private surveillance, online stalking, threats and abuses, and exposure to offensive media content (Buni & Chemaly, 2014). Studies reported that almost half of the women using online dating face abuses (Odongo & Rono, 2017). Mobile devices have also acted as a rising medium for sexual harassment towards women (Hassan, Unwin, & Gardezi, 2018). Issues can often become as critical as leading to gender-based violence through ICTs (Thakur, 2018). The problems take a critical turn in conservative developing countries like India, Pakistan, Israel, and Indonesia, where women identity is defined by the conservative cultural structure of the societies (Barendregt, 2008; Dalal, 2008; Hassan et al., 2018; Lemish & Cohen, 2005; Thakur, 2018). This leads us to suggest a critical lens for the analysis of gender roles of ICT4D.

## 3. LITERATURE

The broader information systems (IS) literature has acknowledged the influence of gender on technology usage (Jackson et al., 2008; Trauth, 2013). The lack of female users and low women participation has dominated the concerns of IS researchers (Dholakia, Dholakia, & Kshetri, 2004). The gender bias extends to the under-representation of women in IT professions and education (Drury, 2011; Major, Morganson, & Bolen, 2013; McGee, 2018). The IT studies with gender focus are not limited to IS literature but extend to gender studies. The examination of such studies shows that there is a persistent influence of IT in widening the gender bias.

ICT4D literature, in general, has highlighted the power of ICTs in the elimination of gender biases, by bridging the divide (Obayelu & Ogunlade, 2006). Several ICT4Ds are specifically designed for the benefit of women beneficiaries (Seth et al., 2020). Community radio has been recognized as a communicative tool for local women through local appropriation, leading to effective participation (Asiedu, 2012). ICTs such as mobile phones and the Internet have been considered as instruments for advancing socioeconomic development, with implications for women concerning empowerment and employment (Danjuma et al., 2015).

Despite these benefits, the prominent gender digital divide constrains the inclusion of women and girls in participation in the ICT4D projects. ICT4D projects often fail to reach the women beneficiaries, though designed to include women empowerment, due to gender-based challenges including the non-usage of mobile phones by women (Potnis, 2014). The freedom to use ICTs become gender-biased significantly when observing the patterns through lenses of culture, religion, and traditions, particularly to the non-Western countries (Ibtasam et al., 2019). Further, ICTs have acted as empowerment only for the elite women, continuing to bar participation of the under-represented, as observed in the #MeToo movement on social media confined to the privileged class of women protesting against sexual harassment (Moitra, Ahmed, & Chandra, 2021; Moitra et al., 2020). These are indications of the gendered nature of ICT (Danjuma et al., 2015; Ibrahim & Adamu, 2016), but the field needs further examination through critical research to examine ICT4D adoption and the barriers for women in terms of low access, social bias, and misuse by sexual abusers (Hassan et al., 2018).

Keeping in mind that the mere introduction of ICT4D projects cannot eradicate the deep social structures of gender bias, we suggest that ICT4D needs a more significant critical stance to analyze its role using a gender-focused lens.

## 4. THEORETICAL BACKGROUND: CRITICAL RESEARCH

The critical research paradigm evaluates and critically analyses the social world and its structures (De' et al., 2017). It deals with "… critiquing existing social systems and revealing any contradictions and conflicts that may inhere within their structures" (Orlikowski & Baroudi, 1991, p. 19). "For more than thirty years of critical research in information systems (IS) has challenged the assumption that technology innovation is inherently desirable and hence to the benefit of all" (McGrath, 2005). With unequal benefits for individuals across gender-based categories as our underlying assumption, critical research provides the appropriate ontology for data analysis (Brooke, 2002). Therefore, in this study, as we aim to examine the gendered nature of ICT4D, critical theory would serve as a tool to challenge the status quo of 'ICT reducing gender divide' and using social theories to reveal the underlying biases or challenges if any. The methodological phases of analysis for critical research in IS have been elaborated by Myers & Klein (2011), which we have followed for the analysis.

| **Principle** | **Principles for Critical Analysis** (De' et al., 2017; Myers & Klein, 2011) | **Critical Analysis Principles for ICT usage by Women** |
|---|---|---|

| Principle-1 | Using core concepts from critical social theorists | Gender Theories like Gender Essentialism, Feminism Theory, Critical Feminism, and Gender schema theory (See the list of Gender theories used in IS studies by Trauth, 2013)), (Sen, 2001; Spivak, 1988) |
|---|---|---|
| Principle-2 | Taking of value position | Strong stand for equal rights of men and women through ICT; Gender-based biases need to be addressed |
| Principle-3 | Challenging prevailing beliefs and social practices | Challenge the notion that ICT empowers women and is free from gender discrimination |
| Principle-4 | Individual emancipation | How is the ICT providing scope of emancipation for women users |
| Principle-5 | Provenience or local history that gives rise to certain social conditions | Understanding the local history and gender-based culture of the society; What is the history of gender inequality in society? |
| Principle-6 | Representation of how ICT shapes the society | Has the diffusion of ICT reduced/enhanced the predominant gender divide in society? Does ICT hold special meaning to women and expression? |
| Principle-7 | Improvements in society | How the socio-political issues that give rise to dark concerns for women using ICTs can be addressed. |

**Table 1. Framework for Critical Analysis of ICT Usage by Women**

## 4.1. Framework for Critical Analysis

The framework for critical analysis of gender studies (Table-1) has been developed using the core principles of IS critical research by (Myers & Klein, 2011) for the stepwise guideline for analysis, with the inclusion of provenience and the structures of the social world (De' et al., 2017). The first principle suggested by Myers & Klein, the identification of concepts by critical social theorists, will require the selection of feminist and gender theories for our framework. The existing theories often have a Western origin while missing the cultural and social issues characteristic of the post-colonial subaltern world (Masiero, 2020). Therefore, we include the seminal theory by Spivak (1988) on the citizens of the postcolonial world and Sen's capability approach (Sen, 2001) to understand freedoms and empowerment of women of the global south embedded in the deep social structures of years of systematic gender-based oppression. When the particular theory has been chosen for the analysis of specific ICTs, the next principles of 'challenging prevailing beliefs' and 'individual emancipation' will be directed by the theoretical stances. Individual emancipation can be established through deep analysis of the participation of women in terms of usage and individual benefits through the ICT. The examination of provenience becomes essential to disentangle the gender issues submerged in traditional layers of oppression, as the local history would reveal the gender-based culture and traditions. The final steps would reveal the impact of the ICT in terms of usage by women beneficiaries and greater societal development.

A critical analysis would reveal both the positive and the darker sides of ICTs and women, bringing out the socio-political issues that give rise to certain disturbing results from an empowering technology (Andalibi et al., 2016; Munyua, Mureithi, & Githaiga, 2010). There has been evidence of gender implications for technostress as a darker side of IT (D'Arcy, Gupta, Tarafdar, & Turel, 2014). Critical research paradigm in information system research suggests six principles that encapsulate the involvement of concepts by social theorists, encourages evaluators to take a value position, and challenge pre-existing societal beliefs for further improvement of the society, through individual emancipation (Myers & Klein, 2011). Additionally, the principle of 'provenience' or the investigating of the local history, and the principle of 'representation' of how the lives of the (women) users are affected by the ICT, offers a deeper and stronger critical lens for the understanding of ICT4D (De' et al., 2017). Additionally, there should be subaltern theoretical considerations, since most influential theories in the previous literature, originating predominantly in the West, often miss out on the nuances of the 'Global South'(Masiero, 2020). Therefore, in this paper, we develop a framework for the evaluation of existing ICTs with the lenses of critical IS research for the developing countries (See Table-1). Reviewing past ICT4D literature and examining the ICTs across the global south, like social media, mobile phones, digital banking, etc., we critically examine its role in women empowerment and the reasons behind the negative concerns from these technologies.

## 5. METHODOLOGY

We conduct a systematic literature review of the past ICT4D studies, but do not confine our search to ICT4D or IS literature. We shortlist papers from various social and psychological domains, with the keyword search with variations of the phrases, 'gender and information technology, 'gender and ICT4D', 'women empowerment and information technology, and 'ICT4D and women'. The purpose is to select the various sub-domains of gender roles of ICT4D and examine them critically using the developed framework.

## 6. PRELIMINARY RESULTS OF CRITICAL ANALYSIS

The following six sub-domains emerged from the preliminary literature review of gender-based ICT4D studies.

### (1) Social Media

Social media has played a significant role in enabling freedom of expression for women across the globe, including the developing countries (Newsom & Lengel, 2012). We will critically analyze social media usage by women, using the framework developed above. In the evaluation, we would investigate how social media has helped women express their views at par with men, but also made them vulnerable to women-focused crimes like online stalking and abuses (Buni & Chemaly, 2014; Newsom & Lengel, 2012). For instance, social media usage failed in the individual emancipation (principle-3) of the underprivileged women who did not raise their voices in the #MeToo movement (Moitra et al., 2020).

### (2) Mobile Phones

Mobile phones are considered one of the most empowering ICTs for the marginalized sectors (Abraham, 2006). Mobile phones have enabled women to convey their romantic expressions through texts in conservative societies (Barendregt, 2008). However, at the same time, mobile phones take the form of an easy weapon for sexual harassment by men (Hassan et al., 2018). The investigation of mobile devices using the framework for critical analysis would reveal the deeper societal issues that give rise to such dark concerns. Mobile phones have also offered freedom to women (Sen, 2001) (Principle-1), and allowed them to express themselves, which is otherwise difficult in the society (Barendregt, 2008) (individual emancipation, principle-4), but also led to sexual abuses causing a similar fear of participation as in the existing conservative society (principle-6 and principle-7).

### (3) Digital Banking

Digital and mobile banking offer socioeconomic development through greater financial inclusion(Donner & Tellez, 2008). The question of inclusion of women in the digital banking ecosystem, in countries like India and Kenya, with a wide gender gap for mobile phone users, is still unanswered. A deeper evaluation of the ICT through critical analysis would show where women stand in terms of financial inclusion and access to various financial services, including digital payment services. When analyzing through critical lenses, mobile banking offers financial freedom (Sen, 2001) (principle-1) to women, but a thorough examination of the literature would reveal the society-level reshaping and improvement (principle-6 and principle-7).

### (4) Telecenters

Although telecenters are installed in countries in less-developed rural locations in Asia, Africa, and Latin America, some reports suggest gender inequality in the usage pattern of these ICTs, that stems from cultural norms and low decision-making power of women (Jorge, 2000; Kumar & Best, 2006;

Lwoga & Lwoga, 2017). A critical lens for analysis of various telecenter projects in developing economies will not only reveal the barriers and constraints faced by potential women users, impacting individual emancipation (principle-4), but also help in understanding the provenience of the culture and its effect on women participation (principle-5).

### (5) Healthcare Information Systems

Women frequently seek health information from online healthcare systems (Warner & Procaccino, 2004). These systems help in both spreading awareness of female-specific ailments and offer online diagnostic support, as well (Bidmon & Terlutter, 2015). It would be interesting to critically evaluate the vast gender differences in the usage patterns of these systems, with women being the more regular users. Social issues like the difficulty of women visiting hospitals and difficulty in traveling could be underlying causes of this phenomenon.

### (6) Internet

A critical analysis of the Internet using the framework would reveal the mass empowerment provided to women. The significant disparity in Internet adoption by the two genders (Dholakia et al., 2004) reveals the rising concerns due to the pre-existing social structures.

In summary, the various ICT4D artifacts that enable socioeconomic development including equality, when examined critically, may expose the dark side of ICT and its prevailing gender-based inadequacies. This paper aims to examine these various sub-domains within ICT4D through the critical theory framework developed.

## 7. CONCLUSION

A critical analysis of the various ICTs and their usage patterns by women helps in understanding the deeper social issues that give effect to the disparity across the genders. While ICTs offer empowerment to women through greater expression and easier access to various services, the wide gender gap and the gender-based offenses on digital platforms, expose the dark side of ICTs. How these issues can be addressed at the grassroots level is understood from the evaluation of the provenience and representation of women in the society through lenses of gender theories, as prescribed by critical theorists.

The critical examination of the ICT4D initiatives and women empowerment are examined through feminism and theoretical lenses that include the nuances of the global south. Often ICTs such as, social media and mobile phones offer freedom to women to express their thoughts and concerns (Sen, 2001), and result in individual emancipation, but may not transform into social improvement due to reinforcement of gender biases already in the historical traditions of the region. As we analyze the various ICTs through a critical framework developed based on De' et al. (2017), Masiero (2020), and Myers & Klein, 2011), we reveal the power and underlying barriers to gender empowerment through ICT4D.